\documentclass{spie}  

\usepackage{amsmath,amsfonts,amssymb}
\usepackage{graphicx}
\usepackage{siunitx}
\usepackage{booktabs}
\usepackage{placeins}
\usepackage{subcaption}
\usepackage[colorlinks=true, allcolors=blue]{hyperref}

\DeclareSIUnit{\electron}{e^-}
\DeclareSIUnit{\pixel}{px}
\DeclareSIUnit{\ph}{ph}
\DeclareSIUnit{\ADU}{ADU}
\newcommand{\okina}{\textquotesingle}

\title{Characterization of IBEX LmAPD detectors at CEA for future low-noise NIR astronomy instruments}

\author[a]{Jean Le Gra\"et}
\author[b]{Vincent Affatato}
\author[a]{Marion Baumann}
\author[a]{Olivier Boulade}
\author[a]{Cyrille Delisle}
\author[b]{Eloy Hernandez}
\author[a]{Christian Ketchazo}
\author[b]{Frederic Lemmel}
\author[a]{Vincent Moreau}
\author[a]{Salima Mouzali}
\author[a]{Thibault Pichon}
\author[a]{Fran\c{c}ois Visticot}
\affil[a]{Universit\'e Paris-Saclay, CEA, CNRS, AIM, F-91191 Gif-sur-Yvette, France}
\affil[b]{European Space Agency, ESTEC, Keplerlaan 1, 2201 AZ, Noordwijk, The Netherlands}

\authorinfo{Send correspondence to Jean Le Gra\"et: E-mail: jean.le.graet@gmail.com}

\begin{document}
\maketitle

\begin{abstract}
Future near-infrared (NIR) astronomy missions dedicated to photon-starved science cases, in particular an all-sky astrometric survey extending the legacy of Gaia into the NIR, as identified within the ESA Voyage~2050 programme, require large-format detectors combining sub-electron effective read noise with very low dark current. Linear-mode avalanche photodiodes (LmAPDs) based on HgCdTe meet this need by multiplying the photo-generated charge by an avalanche gain before the readout stage, thereby reducing the relative contribution of the read noise. We report the first electro-optical characterization of an IBEX detector, a $2048\times2048$, \SI{15}{\micro\meter}-pitch HgCdTe LmAPD array developed by Leonardo with the European Space Agency and operated at \SI{80}{\kelvin} on a dedicated bench at CEA-IRFU. We first discuss the central difficulty of characterizing an APD array, the degeneracy between avalanche gain, quantum efficiency (QE), and conversion gain in the measured response, and we then separate the measurements into two categories. Among the quantities that are directly measurable without assumptions, we report a signal-to-noise ratio for a CDS measurement that exceeds that of a Euclid-type H2RG above \SI{8}{\volt} pixel bias under identical low-flux conditions, a photo-response non-uniformity stable at the $\sim$10\% level, and a quantum efficiency-to-excess-noise ratio (QEFR) of \num{0.40} at \SI{10.5}{\volt}. Among the quantities derived under an explicit set of assumptions, we obtain a conversion gain corresponding to a sense-node capacitance of \SI{27}{\femto\farad}, a quantum efficiency of $46\pm12\%$ at \SI{2.5}{\volt}, and an excess noise factor $F=1.15\pm0.14$ at \SI{10.5}{\volt}. These results establish IBEX as a promising European large-format detector for future ultra-low-flux NIR instruments.
\end{abstract}


\section{INTRODUCTION}
\label{sec:intro}
The Gaia mission has set a new standard for all-sky astrometry, mapping the positions and motions of more than a billion stars~\cite{Prusti-2016, Vallenari-2023}, but its operation at visible wavelengths limits its completeness in regions of high dust extinction, such as the Galactic plane and bulge. Extending astrometric surveys into the near-infrared (NIR) would allow of order ten to twelve billion stars to be surveyed across the Galaxy, including the most obscured regions currently hidden from optical instruments~\cite{Hobbs-2022}, a science objective identified within the European Space Agency (ESA) Voyage~2050 programme as a natural extension of the Gaia legacy. A mission of this kind would operate deep in the photon-starved regime and would rely on the scanning of faint sources, placing stringent demands on the focal-plane detector: sub-electron read noise combined with a large-format, low-dark-current array.

Conventional HgCdTe arrays read out through a source-follower stage are limited by the read noise of the readout chain, typically 10--15 electrons rms in a single correlated double sampling (CDS) frame and reducible to a few electrons only through non-destructive sampling and frame averaging~\cite{Boker-2023, Doyon-2023, Kubik-2026}; there is no straightforward path to sub-electron read noise in such devices. Linear-mode avalanche photodiodes (LmAPDs) overcome this barrier by multiplying the photo-generated charge by an avalanche gain $M$ within the diode, before the read-noise penalty is incurred. Because the read noise $\sigma_{\mathrm{read}}$ is added after multiplication, the effective read noise is reduced to $\sigma_{\mathrm{read}}/M$, and sub-electron values are achievable at moderate gain~\cite{Atkinson-2018a, Finger-2023a}. In HgCdTe the avalanche is dominated by electron impact ionization and is nearly deterministic, so the excess noise factor $F$ that quantifies the multiplication noise remains close to unity, preserving the signal-to-noise ratio in the shot-noise regime.\cite{Rothman-2018}

Leonardo UK has developed a family of HgCdTe LmAPD arrays using metal-organic vapour-phase epitaxy (MOVPE) and a mesa architecture, in which bandgap engineering switches off the junction-related dark-current sources and ensures that the dark current experiences a lower avalanche gain than the photo-signal~\cite{Baker-2023}. The SAPHIRA device ($320\times256$, \SI{24}{\micro\meter} pitch) is used as a wavefront sensor at several observatories and has demonstrated read noise below \SI{0.3}{\electron} rms and single-photon imaging at high gain~\cite{Atkinson-2018a}. The Ike~Pono array ($1024\times1024$ and $2048\times2048$), developed with the University of Hawai\okina i and NASA, was the first large-format LmAPD~\cite{Claveau-2022}. The IBEX detector studied here is a $2048\times2048$ array developed in collaboration with ESA and optimized for dark-current performance, with the goal of providing a fully European large-format NIR detection chain for ultra-low-flux applications. As part of this program, the D\'epartement d'Astrophysique (DAP) at CEA-IRFU is involved in the electro-optical characterization of the device, carried out on benches previously operated for the ALFA/CAGIRE program.

Several papers have already reported on the SAPHIRA and Ike~Pono devices, but IBEX has not yet been characterized in the literature. Our aim here is not to verify the device against a set of specifications but to present to the community the results of our characterization and what can be expected from these detectors. The paper is organized as follows. Section~\ref{sec:detector} describes the detector and its readout circuit, and Sec.~\ref{sec:theory} derives the theoretical expressions for the signal, its noise, and the resulting signal-to-noise ratio of a linear-mode APD. Section~\ref{sec:experiment} presents the experimental setup, namely the characterization bench and the reference-flux measurement. Section~\ref{sec:challenges} discusses the specific difficulty of characterizing an APD array, namely the degeneracy between avalanche gain, quantum efficiency, and conversion gain in the measured response. Section~\ref{sec:direct} then reports the figures of merit that are directly measurable without assumptions, namely the signal-to-noise ratio, the PRNU and pixel outliers, and the QEFR. Finally, Sec.~\ref{sec:derived} derives the conversion gain, the quantum efficiency, and the excess noise factor under an explicit set of assumptions, and Sec.~\ref{sec:conclusion} summarizes the results and discusses the perspectives of this work.

\section{The IBEX detector}
\label{sec:detector}

\begin{figure}[ht]
\begin{center}
\includegraphics[width=0.55\textwidth]{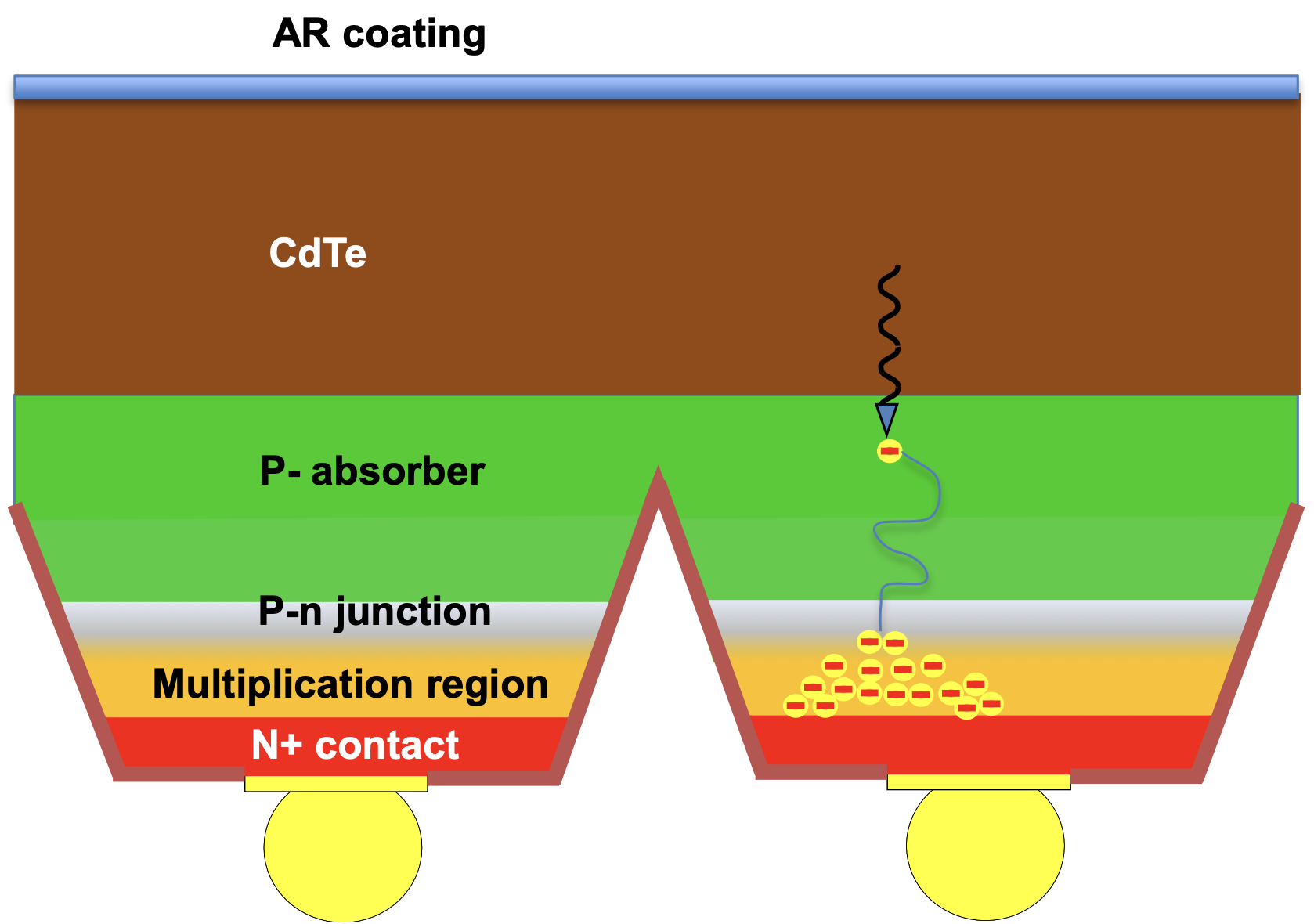}
\end{center}
\caption[pixel]{\label{fig:pixel} Schematic cross-section of an IBEX HgCdTe LmAPD pixel. Photons are absorbed in the wide-bandgap absorber; the photoelectrons drift into a confined narrow-bandgap multiplication region where the avalanche takes place before charge collection and source-follower readout.}
\end{figure}

IBEX is an HgCdTe avalanche photodiode array hybridized to a Leonardo silicon readout integrated circuit (ROIC). Its photosensitive layer, illustrated in Fig.~\ref{fig:pixel}, is very similar to that of the Ike~Pono array detailed in Refs.~\citenum{Claveau-2022, Claveau-2024}, so we describe it only briefly here. Grown by MOVPE, the layer comprises an absorber region with a cut-off wavelength of \SI{2.5}{\micro\meter}, where most of the photons are absorbed, and a multiplication region with a longer cut-off wavelength of \SI{3.5}{\micro\meter}, where the photo-generated electrons are accelerated and multiplied. The multiplication, quantified by the avalanche gain, is set by the pixel bias, defined as the reverse-bias voltage applied across the diode. The difference in cut-off wavelength gives the device a bias-dependent spectral response. Without avalanche, photons between \SI{2.5}{\micro\meter} and \SI{3.5}{\micro\meter} can still be absorbed directly in the multiplication region and contribute to the signal, so the effective cut-off of the detector is \SI{3.5}{\micro\meter}. Under avalanche, however, the electrons generated in the multiplication region are not multiplied, so their relative contribution collapses compared with that of the electrons generated in the absorber and multiplied; the response beyond \SI{2.5}{\micro\meter} then drops sharply and the effective cut-off shifts to \SI{2.5}{\micro\meter}, as measured in Ref.~\citenum{Finger-2023a} for a Saphira detector. The second specificity of the layer is a mesa structure that physically separates the pixels, reducing the crosstalk and confining the photons within each pixel to maximize their absorption.

The array comprises $2048\times2048$ pixels on a \SI{15}{\micro\meter} pitch, complemented by four top and four bottom reference rows. The ROIC is a low-glow design operating from a \SI{3.7}{\volt} supply, the avalanche bias excepted, with a source-follower pixel readout and selectable 4, 8, or 16 analog outputs. The readout is configurable among non-destructive, read-reset-read, and interleaved modes, and multiple windows can be defined for both readout and reset. The detector is operated at \SI{80}{\kelvin} and clocked at a \SI{270}{\kilo\hertz} pixel rate, giving, in the 16-output configuration used here, a frame rate of the order of \SI{1}{\hertz}. Its principal characteristics are summarized in Table~\ref{tab:ibex}.

\begin{table}[ht]
\caption{Principal characteristics of the IBEX detector.}
\label{tab:ibex}
\begin{center}
\begin{tabular}{|l|l|}
\hline
\rule[-1ex]{0pt}{3.5ex} Format & $2048\times2048$ px + 4 top / 4 bottom reference rows \\
\hline
\rule[-1ex]{0pt}{3.5ex} Pixel pitch & \SI{15}{\micro\meter} \\
\hline
\rule[-1ex]{0pt}{3.5ex} Cut-off wavelength & \SI{2.5}{\micro\meter} (avalanche) / \SI{3.5}{\micro\meter} (no avalanche) \\
\hline
\rule[-1ex]{0pt}{3.5ex} Pixel clock & \SI{270}{\kilo\hertz}, 16 outputs, $\sim$\SI{1}{\hertz} frame rate \\
\hline
\rule[-1ex]{0pt}{3.5ex} Pixel readout & Source-follower \\
\hline
\rule[-1ex]{0pt}{3.5ex} Supply voltage & \SI{3.7}{\volt} (excluding avalanche bias) \\
\hline
\rule[-1ex]{0pt}{3.5ex} Operating temperature & \SI{80}{\kelvin} \\
\hline
\rule[-1ex]{0pt}{3.5ex} Power dissipation & $<$ \SI{50}{\milli\watt} \\
\hline
\end{tabular}
\end{center}
\end{table}

\FloatBarrier
\section{Signal and noise in a linear-mode APD}
\label{sec:theory}

Before presenting the measurements, we establish the expressions for the mean signal, its variance, and the resulting signal-to-noise ratio in a linear-mode APD. These expressions underpin the interpretation of the results reported in Sec.~\ref{sec:direct} and define the regime in which avalanche gain is beneficial. In such a device, the photo-generated charge is multiplied by an avalanche gain $M$ before being sensed by the readout stage so that the measured signal and its variance depend on both the statistics of the multiplication process and those of the incoming photons. The multiplication is itself a stochastic process: the
number of secondary carriers produced by a given photoelectron fluctuates from event to event, and this dispersion adds a noise contribution that is absent from a conventional detector. It is quantified by the excess noise factor,
\begin{equation}
\label{eq:enf}
F = 1 + \frac{\sigma_M^2}{\langle M\rangle^2}\, ,
\end{equation}
where $\langle M\rangle$ and $\sigma_M^2$ are the mean and the variance of the gain. A noiseless multiplication would give $F=1$. For HgCdTe APDs, $F$ typically lies in the range $[1,\,1.4]$~\cite{Rothman-2018}. In linear-mode APD, each photon that is absorbed and converted generates a single primary electron, which then triggers its own avalanche and produces $M_i$ carriers at the sense node. The total collected charge is therefore the sum of the individual contributions of all primary electrons generated during the integration, and the signal $S$ measured in ADU can be written as
\begin{equation}
\label{eq:signal-sum}
S = \frac{1}{c_g} \sum_{i=1}^{N_{\mathrm{ph}} \times QE} M_i  ,
\end{equation}
where $c_g$ is the conversion gain in \si{\electron\per\ADU}, $N_{\mathrm{ph}}$ is the number of incident photons per pixel, $M_i$ is the instantaneous gain experienced by the $i$-th primary electron, and $QE$ is the quantum efficiency. Both $N_{\mathrm{ph}}$ and the individual gains $M_i$ are random variables, so that $S$ follows a compound distribution: the number of terms in the sum fluctuates according to the photon statistics, and each term fluctuates according to the statistics of the avalanche. The upper bound of the sum is itself a random variable: the number of photo-generated primary electrons. Its mean is $QE \times \langle N_{\mathrm{ph}} \rangle$ and, as a binomial selection of a Poisson process, it remains Poisson-distributed. Provided that the avalanche gain is independent of the number of absorbed photons, the linearity of the expectation gives the averaged signal as
\begin{equation}
\label{eq:mean-signal}
\langle S \rangle = \frac{QE \times \langle N_{\mathrm{ph}} \rangle \times \langle M \rangle}{c_g} .
\end{equation}
The variance of the signal is obtained by applying the law of total variance to Eq.~(\ref{eq:signal-sum}), conditioning on the number of primary electrons. Three assumptions are required: the multiplications undergone by the individual primary electrons are mutually independent, the gain statistics are the same for all of them and independent of the number of absorbed photons, and the primary electrons obey Poisson statistics. The derivation, given in Appendix~\ref{sec:variance-derivation}, yields
\begin{equation}
\label{eq:var-signal}
\sigma_S^2 = \frac{QE \times \langle N_{\mathrm{ph}} \rangle \times F \times \langle M \rangle^2}{c_g^2}\, .
\end{equation}
The excess noise factor $F$ therefore appears as a direct multiplicative penalty on the signal variance, while leaving the mean signal unchanged.

Equations~(\ref{eq:mean-signal}) and~(\ref{eq:var-signal}) describe the photon contribution alone. In practice two additional noise sources contribute to a measurement: the dark current and the read noise $\sigma_r$ of the readout chain, both expressed here in ADU. The read noise is added after multiplication and is therefore not amplified. The dark current, on the other hand, may be generated at different depths in the structure and does not necessarily experience the same avalanche gain as the photo-generated carriers. We do not attempt to model its statistics and simply denote its contribution to the variance $\sigma_{\mathrm{DC}}^2$, without assuming any particular multiplication factor or noise distribution. The signal-to-noise ratio of a single CDS measurement, for which the read noise contributes twice, is then
\begin{equation}
\label{eq:snr-full}
\mathrm{SNR} = \frac{QE \times \langle N_{\mathrm{ph}} \rangle }{\sqrt{QE \times \langle N_{\mathrm{ph}} \rangle \times F \,  + \dfrac{c_g^2}{\langle M \rangle^2} \left( \sigma_{\mathrm{DC}}^2 + 2\sigma_r^2 \right) }}\, .
\end{equation}
the two terms in the denominator of Eq.~(\ref{eq:snr-full}) show the two different noise regimes of the APD. When looking at high fluxes, the number of incident photons $N_{\mathrm{ph}}$ is high and the second term of the denominator becomes negligible. This is the shot noise-limited regime, and as shown in the equation, in this regime, the APD is useless, as it only decreases the SNR by a factor $\sqrt{F}$. On the other hand, when looking at very low fluxes, the second term is not negligible, and this is where the APD is very useful, as it decreases the contribution of the readout noise by a factor $\langle M \rangle^2$. 

\section{EXPERIMENTAL SETUP}
\label{sec:experiment}

\subsection{Characterization Bench}
\label{sec:bench}

\begin{figure}[ht]
\begin{center}
\begin{subfigure}[b]{0.40\textwidth}
\centering
\includegraphics[height=6.4cm]{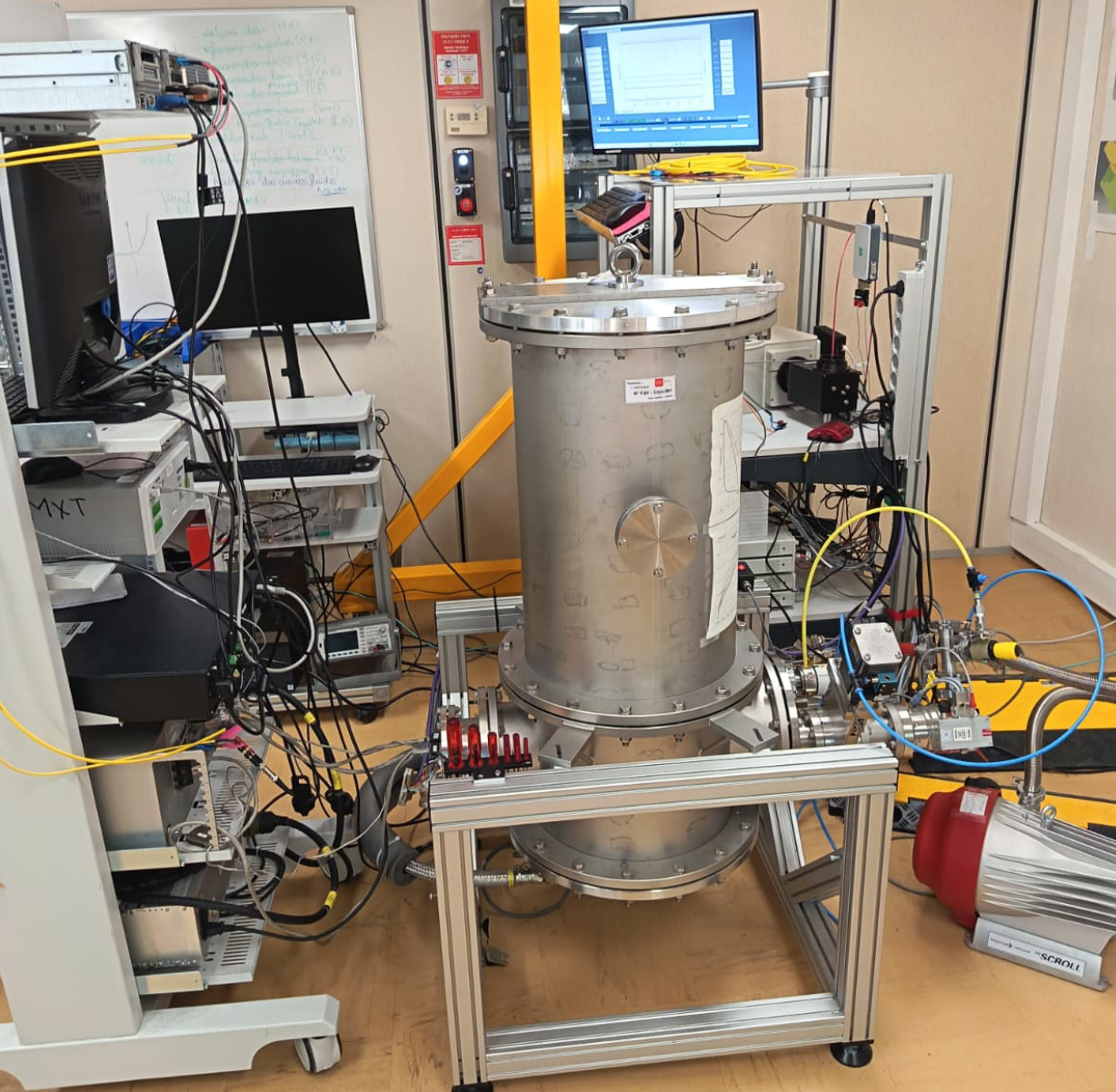}
\caption{Verticalix cryostat on the bench.}
\label{fig:cryo}
\end{subfigure}
\hspace{0.03\textwidth}
\begin{subfigure}[b]{0.40\textwidth}
\centering
\includegraphics[height=6.4cm]{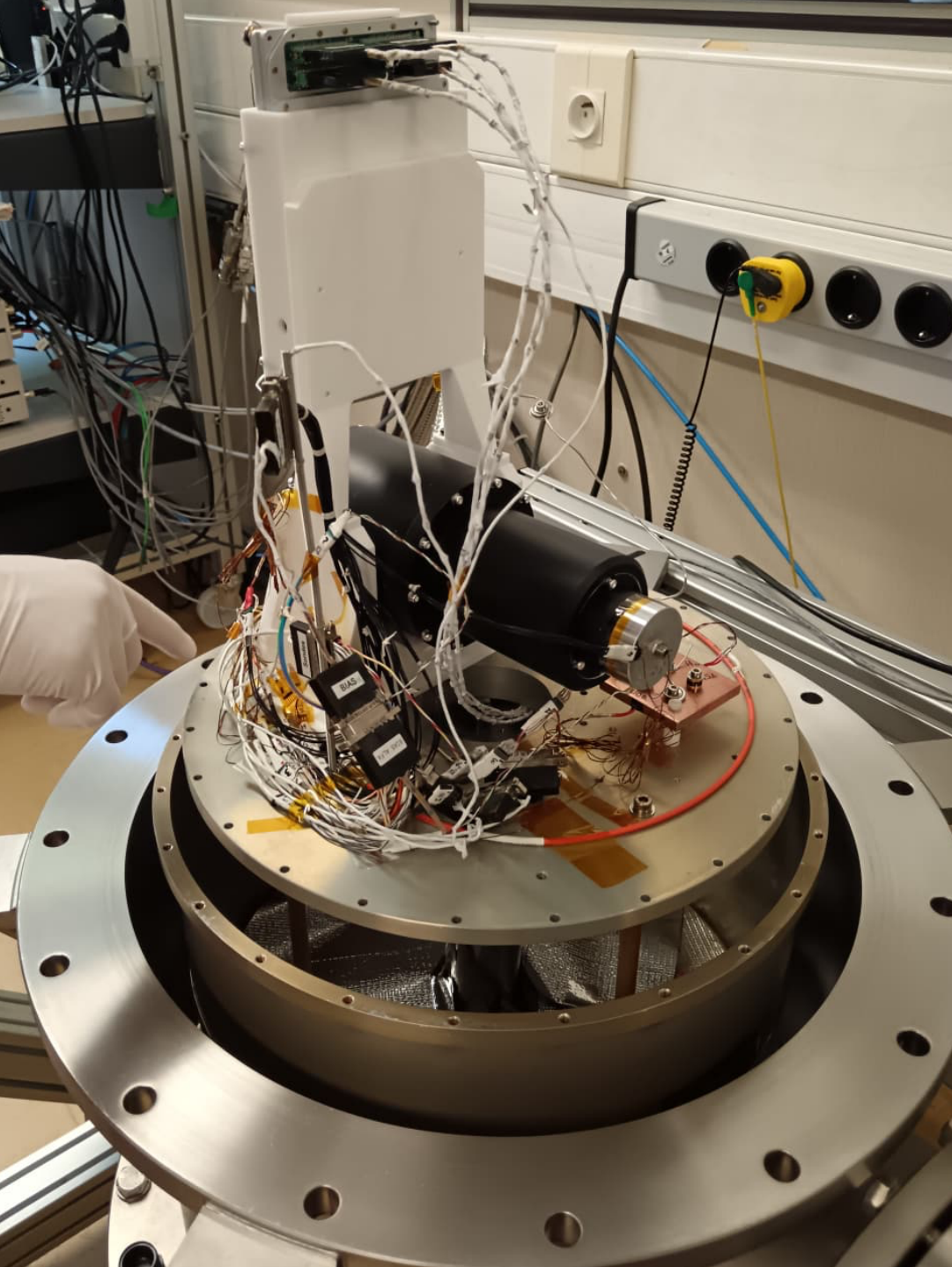}
\caption{Cold stage and detector mount.}
\label{fig:cold_stage}
\end{subfigure}
\end{center}
\caption[bench]{\label{fig:bench} The IBEX characterization bench. (a) The vacuum vessel is mounted on its support frame, with the pumping line and the acquisition and control racks visible on either side; (b) the cold stage is shown before closing the cryostat, with the detector and its mount, the cold preamplifier, the internal illumination assembly, and the cryogenic harness. The array is driven by an ESO NGC controller with an external avalanche-bias supply and read out through the cold preamplifier; illumination is provided either internally (LED / black body) or through an optical fiber.}
\end{figure}

The detector was characterized in the Verticalix cryostat, shown in the left picture of Figure~\ref{fig:bench}, which had previously been used for the characterization of the ALFA detector~\cite{Gravrand-2022}. The cold stage and the detector mechanical assembly inside the cryostat are visible in the right picture of the same figure. IBEX is driven using an ESO New General Detector Controller (NGC) with an external supply delivering the avalanche bias (NGC is limited to 5 V). A cold preamplifier is also used (inside the white mount in the right picture) to amplify the output signal of IBEX and to create the differential signal needed for the NGC. Both the detector and the preamplifier are regulated in temperature using a temperature sensor and a heater driven by a 336 Lakeshore temperature controller (for security needs, the regulation of the detector is redundant). The cryogenic system can cool the detector down to less than \SI{30}{\kelvin}, with temperature fluctuations below \SI{0.7}{\milli\kelvin} over \SI{90}{\min} at the \SI{80}{\kelvin} operating point.
The illumination is flexible: an internal source (LED or black body) can be placed at the end of the optical tube (grey part at the tip of the dark tube housing the detector), or an internal fiber can be plugged into the end of this tube in order to use an external light source. In this work, two light sources were used: an internal OIS-150 1300p LED, driven by a Keithley Source/Measure Unit, and an external QTH (Quartz Tungsten Halogen) lamp combined with a Horiba iHR-320 monochromator. 
To be able to ensure the safety of the detector (which should not operate at room temperature and needs to stay in a clean environment), all the different subsystems are monitored with the NGC computer in order to be able to shut down the detector if needed. A Pfeiffer Omnicontrol was also used to close the different pressure valve and to stop the cooling in case of a pressure leak.

\subsection{Reference-Flux Measurement}
\label{sec:calib}

Measuring the absolute response of the IBEX detector requires precise knowledge of the incident photon flux. This measurement is only possible using a calibrated reference detector. For the IBEX characterization, an MCT detector manufactured by CEA-Leti called the CH1403 was used. Its quantum efficiency has been measured independently at DAP in 2022~\cite{Pichon-2022a} and at ESA in 2015~\cite{Crouzet-2015}, yielding \SI{83}{\percent} and \SI{77}{\percent} respectively at \SI{1550}{\nano\meter}. The two values agree within the \SI{12}{\percent} uncertainty quoted for the DAP measurement. This also constrains any drift of the reference detector: the measured QE did not decrease between the two calibrations, so no aging trend is observed over seven years. We adopt the DAP value, obtained on the same bench and with the same procedure as the present campaign. All the measurements reported here were performed at \SI{1550}{\nano\meter}, the wavelength at which the reference detector is calibrated, so that no spectral extrapolation of the reference QE is required. The illumination is provided by the QTH lamp combined with the Horiba monochromator and then injected into a Thorlabs M35L02 step-index multimode fiber (\SI{1000}{\micro\meter} core diameter, \num{0.39} numerical aperture) going into the Verticalix cryostat. 

\begin{figure}[ht]
\begin{center}
\begin{subfigure}[b]{0.50\textwidth}
\centering
\includegraphics[height=3.5cm]{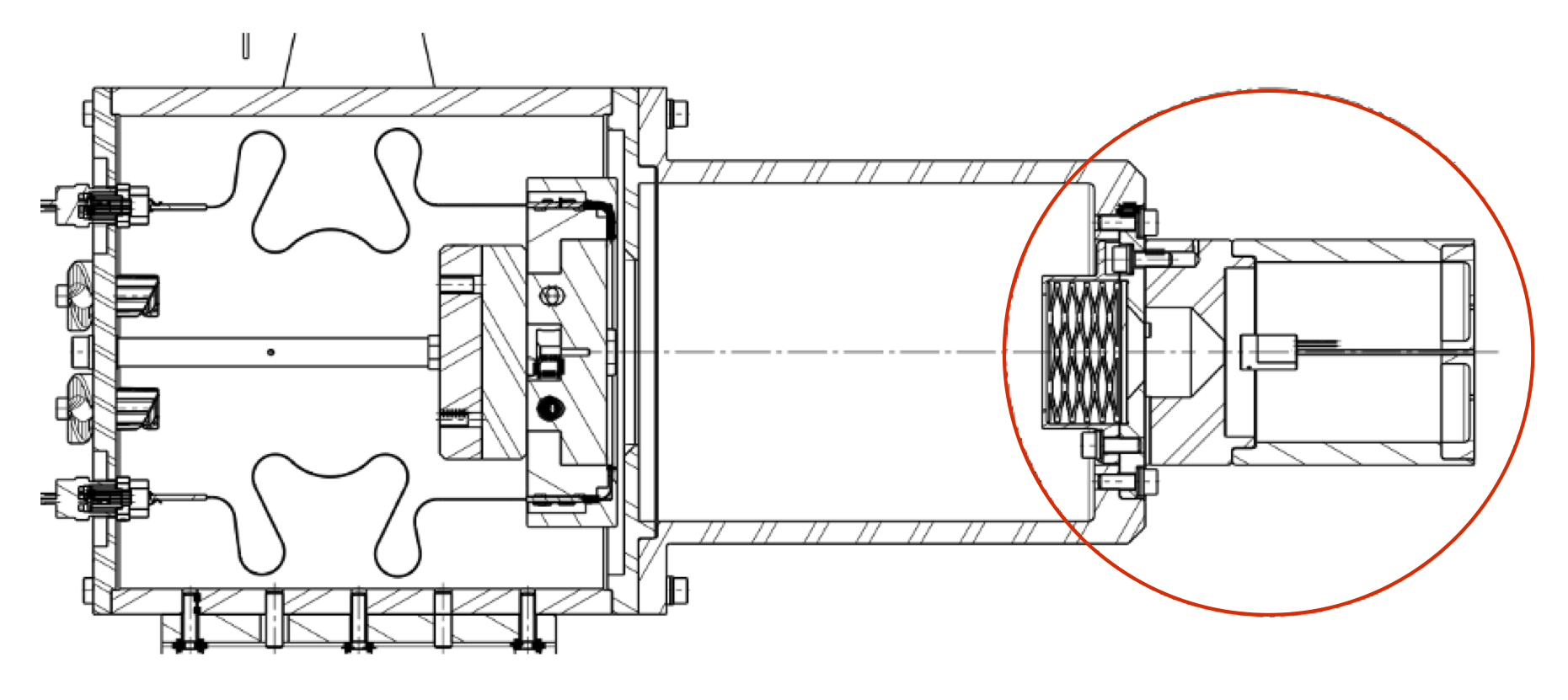}
\caption{IBEX opto-mechanical assembly.}
\label{fig:ibex-assembly}
\end{subfigure}
\hspace{0.01\textwidth}
\begin{subfigure}[b]{0.40\textwidth}
\centering
\includegraphics[height=3.5cm]{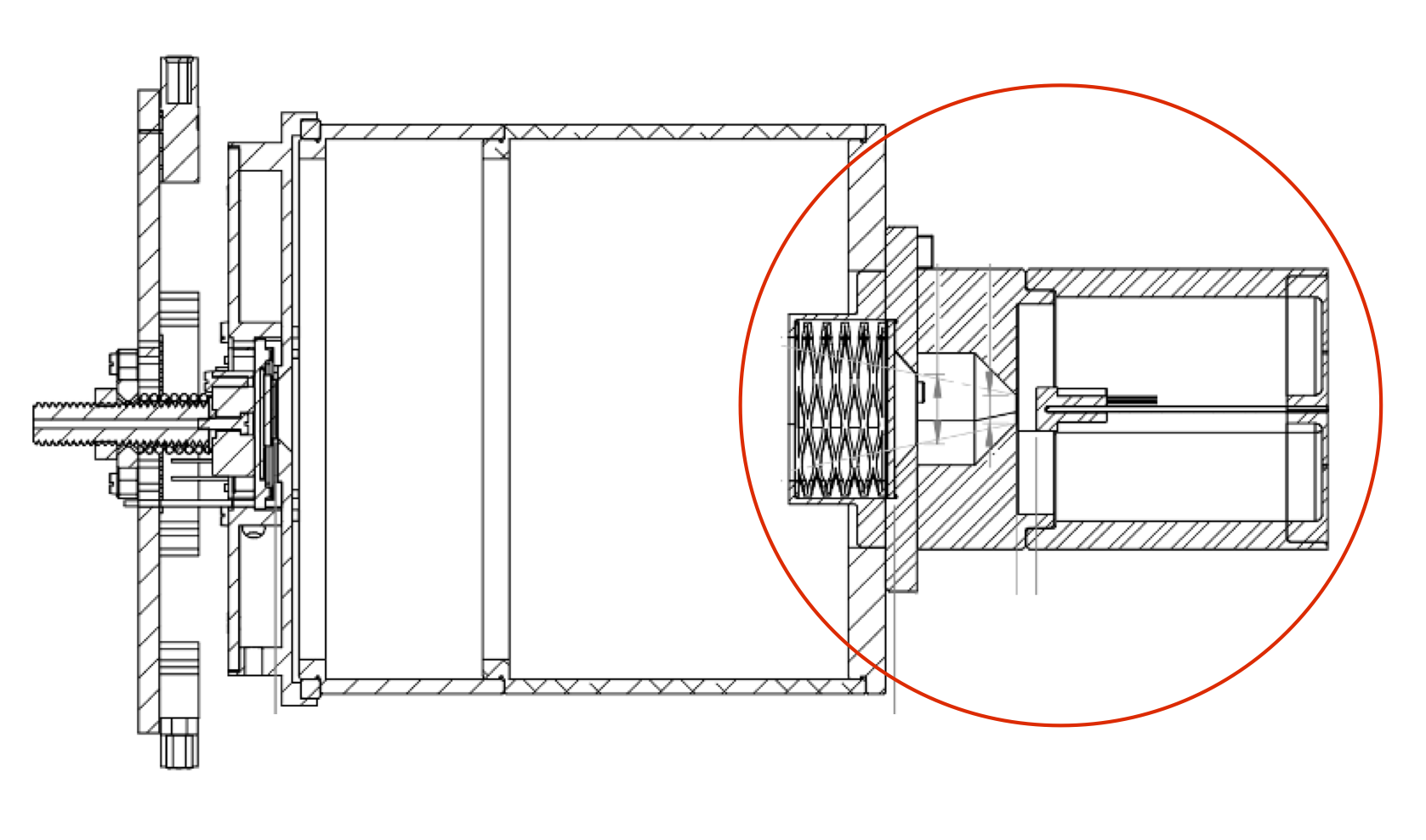}
\caption{CH1403 reference assembly.}
\label{fig:1403-assembly}
\end{subfigure}
\end{center}
\caption[assembly]{\label{fig:assembly} Cross-sections of the opto-mechanical assemblies used for the absolute flux calibration. The optical part circled in red is strictly identical in both assemblies: the two detectors therefore see the same illumination geometry, and the flux measured on the CH1403 is transferred to the IBEX focal plane by a $1/d^2$ scaling with the source-to-detector distance alone.}
\end{figure}

The CH1403 and the IBEX illumination setups share an identical optomechanical geometry, illustrated in Fig.~\ref{fig:assembly}: the optical part circled in red is the same hardware in both configurations, mounted through the same mechanical interfaces. In the configuration used in this study, no filter was installed in the optical mount. The fiber was left strictly untouched between the two measurements, so that its position is rigorously identical in both configurations. The output of the fiber is aligned with the center of both focal planes and is located \SI{2.5}{\milli\meter} from the \SI{4}{\milli\meter}-diameter aperture. The fiber-to-focal-plane distance is \SI{108.3}{\milli\meter} for the CH1403 and \SI{118.1}{\milli\meter} for IBEX. With its \num{0.39} numerical aperture, the fiber illuminates a spot of about \SI{90}{\milli\meter} in diameter at the focal plane. On IBEX, all measurements reported on this work are performed over a $200\times200$ pixel ROI (\SI{3}{\milli\meter} on a side) at the center of the array, well within the flat central part of the illumination profile. The reference flux on the CH1403 is measured over an equivalent central region of the same physical size, ensuring that both detectors sample the same uniform part of the illumination. As the ratio of the fiber core radius to the fiber-to-focal-plane distance is of order $10^{-3}$,
the source can be assumed to be point-like. Then, the absolute photon flux on IBEX can be obtained by applying a $1/d^2$ scaling to the flux measured on the reference detector as
\begin{equation}
\label{eq:fluxtransfer}
\phi_{\mathrm{IBEX}} = \phi_{\mathrm{CH1403}}
\left( \frac{d_{\mathrm{CH1403}}}{d_{\mathrm{IBEX}}} \right)^{2}
= 0.841\,\phi_{\mathrm{CH1403}} ,
\end{equation}
where $d_{\mathrm{CH1403}}$ and $d_{\mathrm{IBEX}}$ are the fiber-to-focal-plane distances given above. The geometrical term in Eq.~(\ref{eq:fluxtransfer}) is only weakly sensitive to the knowledge of the distances: an uncertainty of \SI{0.5}{\milli\meter} on both distances propagates to \SI{1.3}{\percent} on the flux ratio, so that the mechanical tolerances of the mounts are not a limiting factor. The absolute calibration of the CH1403 is therefore the dominant contribution to the systematic uncertainty on the absolute photon flux, and it propagates directly into the quantum efficiency quoted in Sec.~\ref{sec:qe}.

\section{Characterization challenges}
\label{sec:challenges}

Before presenting the results of the IBEX characterization, the main difficulty specific to the characterization of an LmAPD array must be discussed. In a conventional HgCdTe detector, the response is set by the quantum efficiency and the conversion gain alone, and these two quantities can be measured independently. A photon transfer curve (PTC) gives the conversion gain $c_g$ from the slope of the mean-variance curve, which does not depend on the quantum efficiency. The quantum efficiency is then obtained from an absolute flux measurement using $c_g$. The avalanche gain of an LmAPD breaks this separation in two ways, as can be seen from the mean signal and variance derived in Sec.~\ref{sec:theory}.

First, the avalanche adds a third quantity to the response. From Eq.~(\ref{eq:mean-signal}), the mean signal is proportional to $QE \times \langle M \rangle / c_g$, so that any increase of the signal with bias can result from an increase of the quantum efficiency, an increase of the avalanche gain, or a decrease of the conversion gain  The response alone cannot distinguish between these three cases. The situation is made worse by the fact that these quantities vary over the same range of pixel bias. According to Leonardo, the quantum efficiency keeps increasing up to about \SI{6}{\volt}, but we currently have no method to verify this. As a result, neither the exact bias at which the avalanche starts nor the bias above which the quantum efficiency is stable is precisely known.

Second, the avalanche makes the PTC (Photon Transfer Curve) unusable to measure the conversion gain. From Eq.~(\ref{eq:var-signal}), the signal variance is multiplied by $F \langle M \rangle^2$ relative to a conventional detector so that the slope of the mean-variance curve is no longer $1/c_g$ but $F \langle M \rangle / c_g$. A PTC in the avalanche regime, therefore, measures the combination $F \langle M \rangle / c_g$ rather than $c_g$ alone. As we cannot decorrelate $\langle M \rangle$ from QE nor measure the excess noise factor without assuming an a priori model of the avalanche gain, the PTC becomes useless for LmAPD detectors.

\begin{figure}[ht]
\begin{center}
\includegraphics[width=0.9\textwidth]{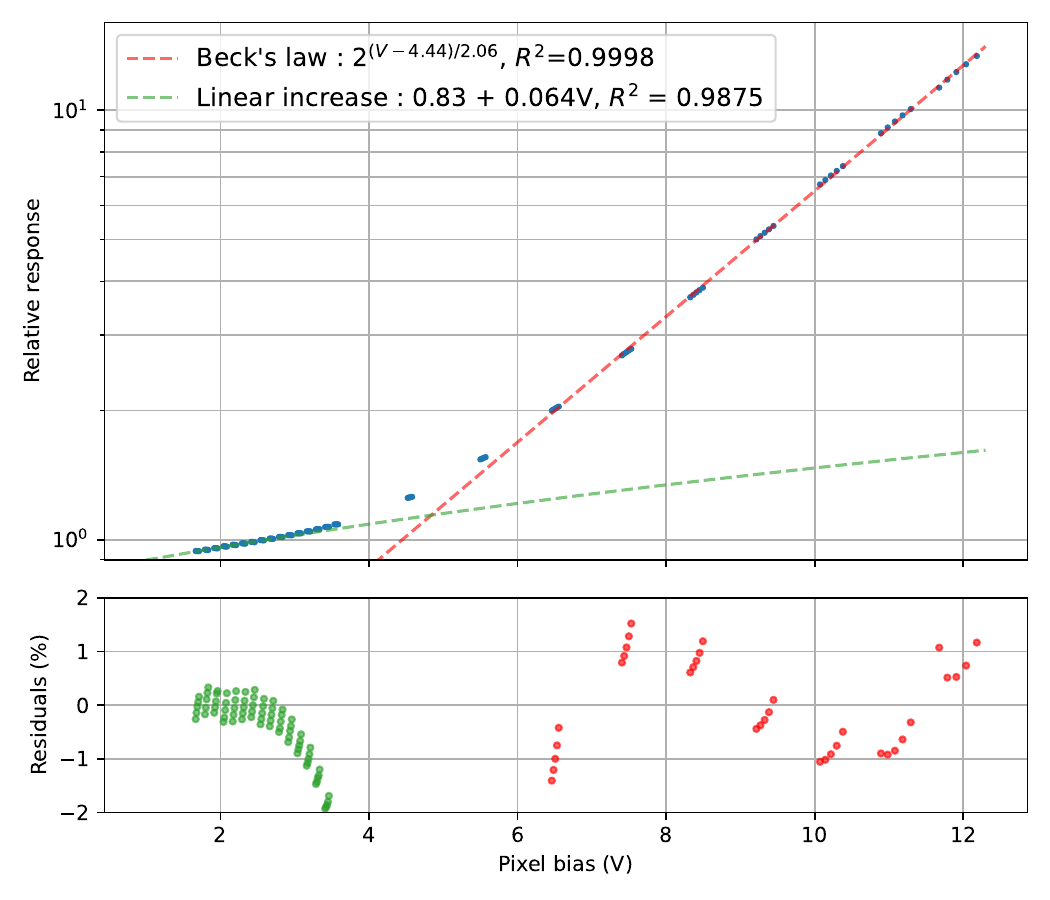}
\end{center}
\caption[response]{\label{fig:response} Relative response $R_{\mathrm{rel}}(V)$ of IBEX as a function of the mean pixel bias, normalized to \SI{2.5}{\volt}. Two regimes are fitted: a linear law $0.83 + 0.064\,V$ below \SI{2.5}{\volt} (green), attributed to a decreasing sense-node capacitance, and Beck's exponential avalanche law $2^{(V-4.44)/2.06}$ above $\sim$\SI{7}{\volt} (red). The bottom panel shows the fit residuals in percent. In the transition region between the two regimes, the avalanche gain, quantum efficiency, and conversion gain vary simultaneously and cannot be disentangled from the response alone.}
\end{figure}

To illustrate these entanglements, we examine how the response of the IBEX detector changes when the pixel bias is increased, the pixel bias being the potential difference applied across the pixel. The response is normalized to its value at the reference bias of \SI{2.5}{\volt}, a choice justified below. The relative response, shown in Fig.~\ref{fig:response}, is obtained from 100 ramps of 10 frames, acquired at \SI{80}{\kelvin}, at a constant photon flux ($\sim$\SI{9000}{\ph\per\second}), a fixed integration time, and an increasing pixel bias. From each ramp, 5 CDS measurements were extracted and averaged across the 100 ramps at each pixel bias. The median of these averaged values over the same $200\times200$ pixel region as in Sec.~\ref{sec:calib}, normalized to its value at \SI{2.5}{\volt}, was then plotted against the mean pixel bias during the acquisition, computed as
\begin{equation}
\label{eq:mean_bias}
\langle V_{px} \rangle = V^0_{px} - \dfrac{1}{2}\,S_{mes}\,\dfrac{e}{C_n} ,
\end{equation}
with $V^0_{px}$ the pixel bias applied at the reset in \si{\volt}, $S_{mes}$ the measured signal in electrons, $e$ the elementary charge in \si{\coulomb}, and $C_n$ the node capacitance in \si{\farad}. The signal in electrons is obtained from the CDS signal in ADU through the conversion gain $c_g$, and both $C_n$ and $c_g$ are taken from the measurement of Sec.~\ref{sec:cg}, whose values are in excellent agreement with those expected by Leonardo. The factor $\tfrac{1}{2}$ assumes that the charge is integrated linearly along the ramp, so that the average charge accumulated on the node is half the total signal $S_{mes}$, and the mean bias is the reset bias reduced by half the total voltage drop. Plotting the response against the mean bias, rather than the reset bias, is a direct consequence of the source-follower readout. The bias is set at the reset, but as photo-generated charges accumulate on the sense node during the acquisition, the node potential decreases, and so does the effective bias across the pixel. The bias therefore droops continuously along the ramp, from its reset value $V^0_{px}$ down to a lower value at the end of the integration, and the avalanche gain, which depends exponentially on the bias, droops with it. Using the reset bias would attribute the measured response to a bias that most of the integrated charge never experiences.

The relative response reveals three regimes. Below \SI{2.5}{\volt}, it increases linearly with pixel bias, and the data are well described by a linear law ($R^2 = 0.9875$). In this range the avalanche is assumed not to have started and the avalanche gain to be unity, so that the increase can be attributed to a decrease of the sense-node capacitance, that is of the conversion gain. As noted above, it could also reflect an increase of the quantum efficiency or a combination of both, but the quality of the linear behavior strongly supports the assumption that the avalanche has not yet started. This is why the relative response is normalized to unity at \SI{2.5}{\volt}, taken as the reference point below which no multiplication occurs and above which Leonardo expects the sense-node capacitance to be stable. Above $\sim$\SI{7}{\volt}, the response increases exponentially and is well described by Beck's law, $M(V)=2^{(V-V_{\mathrm{th}})/V_d}$, with $V_{\mathrm{th}} = \SI{4.44}{\volt}$ and $V_d = \SI{2.06}{\volt}$ ($R^2 = 0.998$), consistent with an avalanche-dominated regime~\cite{Beck-2006}. Between these two regimes lies a transition region where the avalanche is starting while the quantum efficiency is still increasing so that all three quantities vary simultaneously and the response cannot be assigned to any single one of them. Leonardo expects the quantum efficiency to become stable above \SI{6}{\volt}, but we currently have no means of verifying this, since the avalanche gain and the quantum efficiency rise together across the transition and no measurement can isolate one from the other at \SI{6}{\volt}. This behavior is the direct illustration of the degeneracy discussed above. Because none of $QE$, $\langle M \rangle$, or $c_g$ can be extracted from the relative response alone, we adopt a two-step strategy for the rest of the paper. We first report, in Sec.~\ref{sec:direct}, the figures of merit that are directly accessible from the data and require no assumption on the underlying quantities, namely the signal-to-noise ratio, the PRNU and pixel outliers, and the QEFR. We then derive, in Sec.~\ref{sec:derived}, the quantities that require an explicit set of assumptions to be extracted individually: the conversion gain, the quantum efficiency and the excess noise factor.

\FloatBarrier

\section{FIGURES OF MERIT MEASURED WITHOUT ASSUMPTIONS}
\label{sec:direct}
We first report the quantities that can be measured directly, without assuming a value for QE, $c_g$, or $M$. All measurements are performed over the same region of interest (ROI): a $200\times200$ pixel square placed where the fiber illumination is most uniform. The detector is powered on at \SI{150}{\kelvin} and then cooled to its \SI{80}{\kelvin} operating point. Unless otherwise stated, statistics are computed over 100 CDS acquisitions at each operating point.

\subsection{Signal-to-Noise Ratio versus Bias}
\label{sec:snr}
\begin{figure}[ht]
\begin{center}
\includegraphics[width=0.75\textwidth]{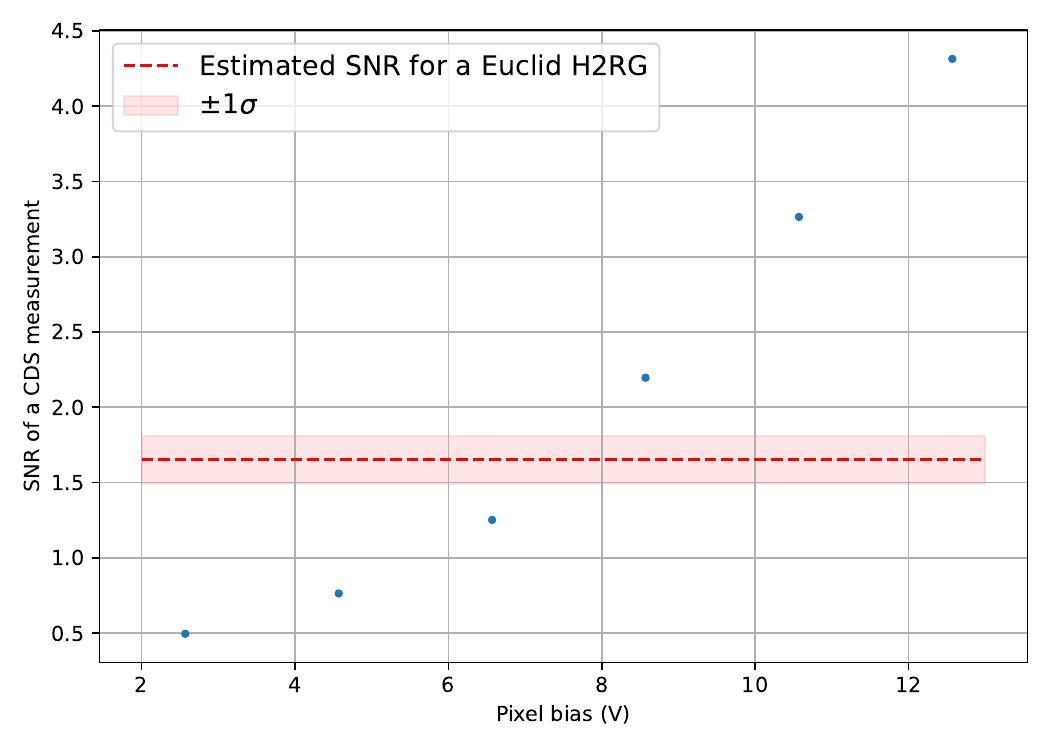}
\end{center}
\caption[snr]{\label{fig:snr} Median SNR over the ROI as a function of pixel bias [Eq.~(\ref{eq:snr})], measured at a constant photon flux of \SI{35}{\ph\per\second} over 100 CDS acquisitions per bias point. The IBEX SNR is compared with the estimated SNR of a Euclid-type H2RG ($\pm1\sigma$ band) under identical low-flux conditions. Above \SI{8}{\volt} IBEX outperforms the H2RG.}
\end{figure}

The signal-to-noise ratio is the figure of merit that ultimately matters to the user, since it directly sets the detection limit of the instrument. It is also directly measurable, without any assumption on $QE$, $\langle M \rangle$, or $c_g$. The SNR is defined as the useful signal divided by the total noise of the measurement,
\begin{equation}
\label{eq:snr}
\mathrm{SNR} = \frac{\langle S\rangle - \mathrm{DC}}{\sigma_S}\, ,
\end{equation}
where $\langle S \rangle$ is the mean signal in ADU, $\mathrm{DC}$ the dark contribution in ADU, and $\sigma_S$ the temporal standard deviation of the signal in ADU. As shown in Sec.~\ref{sec:theory}, the avalanche increases the SNR only in the read-noise-limited regime. To measure it in conditions representative of the intended use, we use a constant photon flux of about \SI{35}{\ph\per\second} and a fixed integration time, and step the pixel bias while measuring the SNR from 100 CDS acquisitions at each bias. The evolution of the median SNR over the ROI with pixel bias is shown in Fig.~\ref{fig:snr}. Because the 100 CDS acquisitions at each bias are needed to estimate the SNR itself, no error bar can be attached to the individual points. 
The SNR increases with pixel bias as expected, from \num{0.5} at \SI{2.5}{\volt} to more than \num{4} at \SI{12.5}{\volt}. For a photon-starved survey such as GaiaNIR, this improvement pushes the detection limit fainter and directly increases the number of stars the mission can measure.
Figure~\ref{fig:snr} also shows the estimated SNR of a Teledyne H2RG~\cite{Blank-2011} with performance similar to that of the Euclid H2RGs~\cite{Kubik-2026}, computed for a CDS measurement of a conventional CMOS detector using
\begin{equation}
\label{eq:euclid_snr}
\mathrm{SNR} = \frac{QE \times \langle N_{\mathrm{ph}} \rangle}{\sqrt{QE \times \langle N_{\mathrm{ph}} \rangle + c_g^2 \left( \sigma_{\mathrm{DC}}^2 + 2\sigma_r^2 \right) }}\, ,
\end{equation}
where $\sigma_{\mathrm{DC}}$ and $\sigma_r$ are the dark and read noise in ADU, $c_g$ is the conversion gain in \si{\electron\per\ADU}, and $QE \times \langle N_{\mathrm{ph}} \rangle$ is the number of photo-generated electrons. The error bar of this estimation comes from the uncertainty in the QE of the reference detector used to obtain the incoming photon flux.
Above \SI{8}{\volt}, the IBEX SNR exceeds that of the Euclid H2RG under identical low-flux conditions. It means that at a single CDS readout, IBEX outperforms the reference near-infrared detector of current space missions in the very-low-flux regime, establishing IBEX as a credible European alternative to HxRG detectors for photon-starved applications. The impact of non-destructive readout schemes, such as Fowler or up-the-ramp sampling, on this comparison remains to be studied.

\subsection{PRNU and Pixel Outliers}
\label{sec:prnu}
\begin{figure}[ht]
\begin{center}
\includegraphics[width=0.85\textwidth]{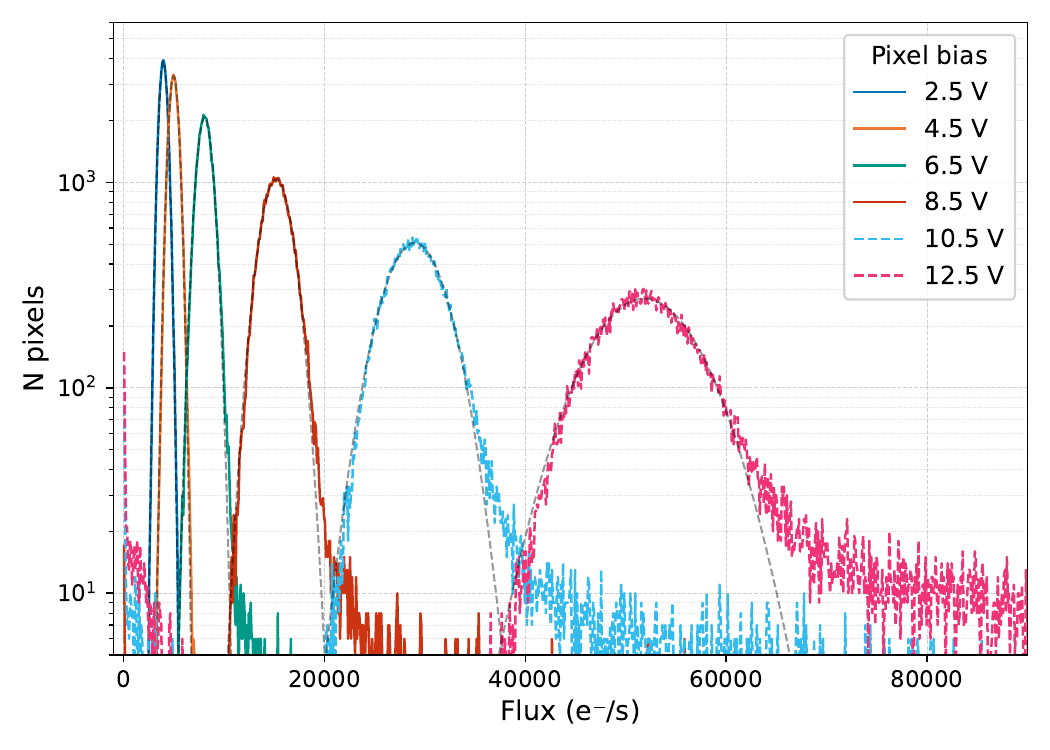}
\end{center}
\caption[prnu]{\label{fig:prnu} Distribution of the measured CDS flux (in \si{\electron\per\second}) over the ROI, estimated across 100 measurements, at different pixel biases for a photon flux of \SI{8600}{\ph\per\second}. The distribution width, which sets the PRNU, remains stable with bias while the tails, and hence the number of $3\sigma$ outliers, grow with increasing bias. At the two highest biases a secondary population of pixels appears near zero flux.}
\end{figure}

The signal-to-noise ratio characterizes the temporal behavior of a pixel, but a large-format detector must also be uniform across the array. A high median SNR is not sufficient on its own: if a large fraction of the pixels behave as outliers, the operability of the detector collapses and the number of observable targets drops accordingly. The avalanche gain varies from pixel to pixel, and this dispersion translates into a spatial non-uniformity of the response. We therefore characterize the photo-response non-uniformity (PRNU) and the fraction of outlier pixels as a function of pixel bias.
The PRNU is computed locally over the ROI: for each $10\times10$ pixel sub-window, the spatial dispersion $\sigma_S / \langle S \rangle$ is evaluated, where $\langle S \rangle$ and $\sigma_S$ are the mean and standard deviation of the signal within the sub-window, and the PRNU is taken as the median over all sub-windows. The measurement is performed from 100 CDS acquisitions at each pixel bias, for a photon flux of \SI{8600}{\ph\per\second}, and both the PRNU and the outlier fraction are equivalent at the lower flux of \SI{35}{\ph\per\second} used for the SNR measurement. Figure~\ref{fig:prnu} shows the corresponding flux distributions at several biases, and Table~\ref{tab:prnu} lists the PRNU and the fraction of $3\sigma$ outliers versus bias.

The PRNU remains nearly constant, at the $\sim$10\% level across the whole bias range. This is higher than expected for this type of device. Leonardo has already identified and addressed the cause of this spatial variation, so the follow-up IBEX detectors are expected to show a lower PRNU than the one measured here. The fraction of $3\sigma$ outliers grows with bias, from 0.7\% at \SI{2.5}{\volt} to 11.8\% at \SI{12.5}{\volt}. 
At the two highest biases (\SI{10.5}{\volt} and \SI{12.5}{\volt}), Fig.~\ref{fig:prnu} shows a population of pixels whose CDS signal falls close to zero, which contributes to the outlier count. The origin of this effect is still under investigation, and this study should help understand where the outliers come from and reduce their number in the next versions of IBEX. Even at \SI{12.5}{\volt}, however, fewer than 12\% of the pixels are outliers, which suggests that the operability of IBEX remains acceptable across the full bias range. A quantitative definition of the operability and the study of its evolution with pixel bias are ongoing.

\begin{table}[ht]
\caption{PRNU and fraction of $3\sigma$ outliers versus pixel bias, at a photon flux of \SI{8600}{\ph\per\second}.}
\label{tab:prnu}
\begin{center}
\begin{tabular}{|c|c|c|}
\hline
\rule[-1ex]{0pt}{3.5ex} Pixel bias & PRNU & Outliers ($3\sigma$) \\
\hline
\rule[-1ex]{0pt}{3.5ex} \SI{2.5}{\volt}  & 9.7\%  & 0.7\%  \\
\hline
\rule[-1ex]{0pt}{3.5ex} \SI{4.5}{\volt}  & 9.5\%  & 1.2\%  \\
\hline
\rule[-1ex]{0pt}{3.5ex} \SI{6.5}{\volt}  & 9.4\%  & 2.6\%  \\
\hline
\rule[-1ex]{0pt}{3.5ex} \SI{8.5}{\volt}  & 9.7\%  & 5.0\%  \\
\hline
\rule[-1ex]{0pt}{3.5ex} \SI{10.5}{\volt} & 10.1\% & 8.2\%  \\
\hline
\rule[-1ex]{0pt}{3.5ex} \SI{12.5}{\volt} & 10.4\% & 11.8\% \\
\hline
\end{tabular}
\end{center}
\end{table}

\subsection{Quantum Efficiency-to-Excess-Noise Ratio}
\label{sec:qefr}

As discussed above, the avalanche gain, the quantum efficiency, and the conversion gain cannot be measured independently without additional assumptions, so the figures of merit that rely on them individually are not directly accessible for an APD. To overcome this limitation, Rothman~\cite{Rothman-2025} introduced a directly measurable, assumption-free figure of merit for APDs, the quantum efficiency-to-excess-noise ratio (QEFR). It is derived from the expression of the SNR in the shot-noise regime. Although this is not the regime in which IBEX is intended to operate, the QEFR provides a direct measure of the quality of the photosensitive layer and allows different APD detectors to be compared on a common basis.

\begin{figure}[ht]
\begin{center}
\includegraphics[width=0.8\textwidth]{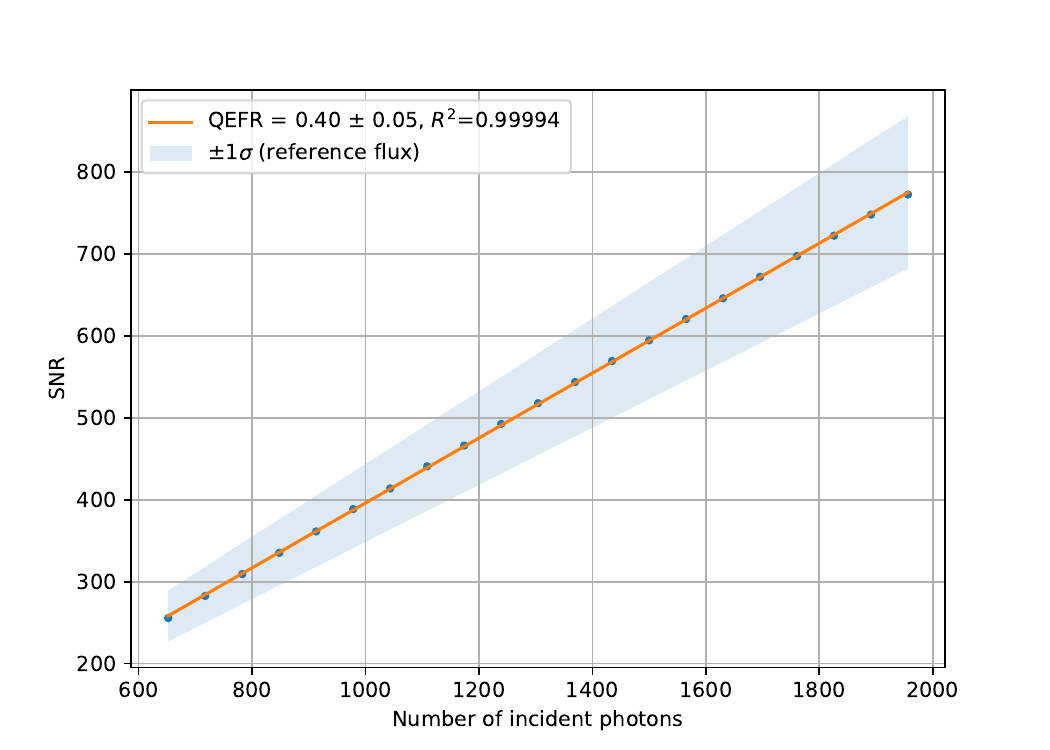}
\end{center}
\caption[qefr]{\label{fig:qefr} Squared SNR over the ROI as a function of the number of incident photons $\langle N_{\mathrm{ph}} \rangle$, at a pixel bias of \SI{10.5}{\volt}. Following Eq.~(\ref{eq:qefr}), the slope of the linear fit gives QEFR $= 0.40 \pm 0.05$ at this bias, the uncertainty being dominated by the systematic uncertainty on the number of incident photons.}
\end{figure}

Starting from the SNR of an APD detector given in Eq.~(\ref{eq:snr-full}), we define the regime in which the second term of the denominator is negligible compared to the first,
\begin{equation}
\label{eq:shot}
\dfrac{c_g^2}{\langle M \rangle^2} \left( \sigma_{\mathrm{DC}}^2 + 2\sigma_r^2 \right) \, \ll \, QE \times \langle N_{\mathrm{ph}} \rangle \times F \, .
\end{equation}
We refer to this as the shot-noise regime, keeping in mind that the avalanche extends it to lower fluxes than for a conventional detector: the factor $\langle M \rangle^2$ in the denominator reduces the contribution of the readout noise term, so that the shot noise dominates down to fluxes at which a detector without avalanche would already be readout-noise limited. In this regime the squared SNR reduces to
\begin{equation}
\label{eq:qefr}
\mathrm{SNR}^2 \simeq \frac{\mathrm{QE}}{F}\langle N_{\mathrm{ph}} \rangle = \mathrm{QEFR} \, \langle N_{\mathrm{ph}} \rangle\, ,
\end{equation}
with $\mathrm{QEFR} = \mathrm{QE}/F$. Since the squared SNR of the incident photon number is $\mathrm{SNR}_{\mathrm{in}}^2 = \langle N_{\mathrm{ph}} \rangle$, the QEFR is the ratio of the detector SNR to the ideal photon-limited SNR, and therefore quantifies the information lost in the detection. For a detector without avalanche, or with a perfectly deterministic one ($F=1$), it reduces to the quantum efficiency. Equation~(\ref{eq:qefr}) shows that the QEFR is obtained, without any assumption, as the slope of the squared SNR plotted against the mean number of incident photons. To measure the IBEX QEFR, 100 CDS measurements at a constant photon flux of about \SI{650}{\ph\per\second} were used, with the integration time increased from \SI{1}{\second} to \SI{3}{\second}, at a pixel bias of \SI{10.5}{\volt}. The SNR is estimated from the 100 measurements at each integration time and plotted in Fig.~\ref{fig:qefr}.

A linear fit of the curve yields QEFR $= 0.40 \pm 0.05$ at this bias, where the uncertainty is dominated by the \SI{12}{\percent} systematic uncertainty on the number of incident photons discussed in Sec.~\ref{sec:calib}. A QEFR of \num{0.40} means that IBEX preserves 40\% of the information carried by the incident photons in the shot-noise regime. This information loss must, however, be put in the context of the regime in which these detectors are used. IBEX is not intended to operate in the shot-noise regime, where the avalanche brings no benefit, but at the very low fluxes where the readout noise dominates. In that regime, the avalanche gain raises the SNR well above the value obtained without avalanche, as shown in Sec.~\ref{sec:snr}, turning sub-threshold sources into detectable ones. The information loss measured by the QEFR is therefore largely outweighed, for a mission such as GaiaNIR, by the large number of new targets made accessible by the avalanche-driven increase in sensitivity. Because it is directly measurable and free of any assumption on the avalanche gain, the quantum efficiency, or the conversion gain, the QEFR is well suited to compare the intrinsic quality of APD photosensitive layers across manufacturers and formats, and we believe it could become a reference metric for this purpose. As a first illustration, Rothman~\cite{Rothman-2025} reports a QEFR of \num{0.45} on a single-element APD. The IBEX value is remarkably close to this figure while being obtained on a full $2048\times2048$ array, which shows that the quality of the photosensitive layer is preserved across a large-format device and remains comparable to that of a single diode.

\section{FIGURES OF MERIT DERIVED UNDER ASSUMPTIONS}
\label{sec:derived}
The figures of merit reported so far characterize the detector without separating the quantum efficiency, the avalanche gain, and the conversion gain. To access these quantities individually, an explicit set of assumptions is required, and we specify them for each derived quantity. These assumptions are based both on the observations made during the characterization and on the expected behavior of IBEX as described by Leonardo.

\subsection{Conversion Gain at \SI{2.5}{\volt}}
\label{sec:cg}

\begin{figure}[ht]
\begin{center}
\includegraphics[width=0.75\textwidth]{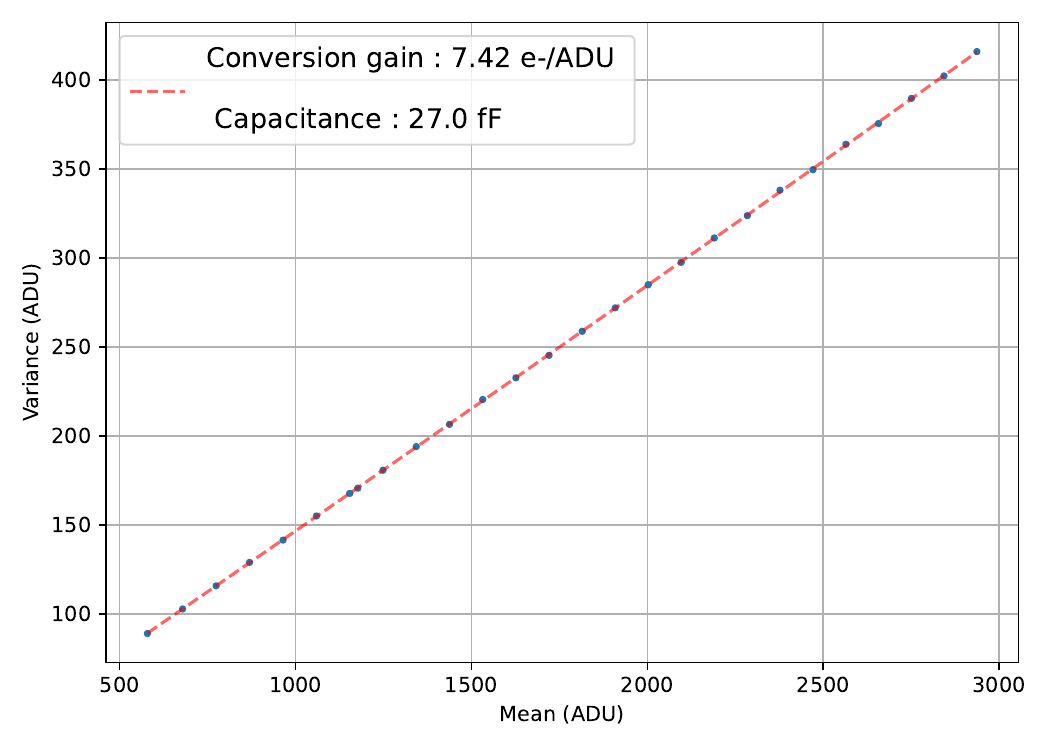}
\end{center}
\caption[ptc]{\label{fig:ptc} Median photon transfer curve (signal variance $\sigma_S^2$ versus mean signal $\langle S \rangle$) over the ROI at a pixel bias of \SI{2.5}{\volt}, obtained from 26 integration times between \SI{2}{\second} and \SI{10}{\second} at constant flux, with an order-2 orthogonal-polynomial fit. The extracted conversion gain of \SI{7.42}{\electron\per\ADU} corresponds to a sense-node capacitance of \SI{27}{\femto\farad}, consistent with the \SI{27.3}{\femto\farad} expected by Leonardo.}
\end{figure}

As shown in Sec.~\ref{sec:challenges}, a mean-variance curve can only be used to measure the conversion gain outside the avalanche regime, where the variance is not multiplied by the excess noise factor and the avalanche gain. We therefore measure it at \SI{2.5}{\volt}, assuming the avalanche gain and the excess noise factor to be unity at this bias. According to Leonardo, the depletion region reaches its maximum width at this bias, so the conversion gain is expected to remain constant above \SI{2.5}{\volt}. The signal variance $\sigma_S^2$ is measured as a function of the mean signal $\langle S \rangle$ using 26 integration times from \SI{2}{\second} to \SI{10}{\second} at a constant flux of $\sim$\SI{8600}{\ph\per\second}, with 100 CDS acquisitions per point. The median mean-variance curve over the ROI is shown in Fig.~\ref{fig:ptc}. The conversion gain is extracted from the linear term of an order-2 orthogonal-polynomial fit, following the method described in Le Gra\"et et al.\cite{Graet-2026, Graet-2024}. As we do not yet have a robust measurement of the inter-pixel capacitance (IPC) of these detectors, no IPC correction is applied. We measure a conversion gain of \SI{7.42}{\electron\per\ADU}, corresponding to a sense-node capacitance of \SI{27}{\femto\farad}. This value is in good agreement with the \SI{27.3}{\femto\farad} expected by Leonardo, although the IPC correction is expected to lower it slightly. The result confirms that the avalanche gain is unity at this bias and validates the choice of \SI{2.5}{\volt} as the reference point.

\subsection{Quantum Efficiency at 2.5\,V}
\label{sec:qe}

Now that the conversion gain is known, the quantum efficiency can be extracted at the same bias of \SI{2.5}{\volt}. Since the avalanche gain is unity there, the mean signal reduces to $\langle S \rangle = QE \times \langle N_{\mathrm{ph}} \rangle / c_g$, so that the quantum efficiency is the only remaining unknown once the conversion gain has been measured and the incident photon flux is known from the reference-flux calibration. It is obtained over the ROI as the mean of 100 CDS measurements, using the conversion gain of Sec.~\ref{sec:cg} to express the signal in electrons and the same reference-flux measurement as in Sec.~\ref{sec:calib} to obtain the incident photon flux. At \SI{2.5}{\volt}, the median quantum efficiency of IBEX over the ROI is
\begin{equation}
\label{eq:qe}
\mathrm{QE} = 46 \pm 12\%\, ,
\end{equation}
where the uncertainty is dominated by the absolute calibration of the reference detector discussed in Sec.~\ref{sec:calib}. This value is lower than expected by Leonardo, which we attribute to two causes. First, Leonardo expects the quantum efficiency to keep increasing up to a pixel bias of \SI{6}{\volt}, whereas our value is measured at \SI{2.5}{\volt}. As discussed in Sec.~\ref{sec:challenges}, we currently have no method to separate the quantum efficiency from the avalanche gain above this bias, so the value at \SI{6}{\volt} cannot be measured directly. Second, the IBEX detector characterized at CEA is suspected to have a defect in its sensitive layer, particularly in the depletion region, which would lower its quantum efficiency at low bias compared with the three other IBEX detectors characterized at ESTEC. We plan to repeat these measurements on another IBEX detector to confirm this. The quantum-efficiency map is also uniform across the ROI and the pixel distribution is Gaussian with a mean value of 46\% and a standard deviation of 5\%, confirming that the reduction affects all pixels uniformly, consistent with a global effect rather than a local one. Even though the measured quantum efficiency is lower than expected, the SNR study of Sec.~\ref{sec:snr} confirms the usefulness of IBEX for very-low-flux applications.

\subsection{Excess Noise Factor at \SI{10.5}{\volt}}
\label{sec:enf}

Another quantity that can be derived is the excess noise factor $F$, which quantifies the noise added by the stochastic nature of the avalanche and was defined in Sec.~\ref{sec:theory} as $F = 1 + \sigma_M^2 / \langle M \rangle^2$. Having measured the quantum efficiency, we can now access $F$, since the two are linked through the QEFR. It cannot be measured directly with our setup, but it is recovered by combining the quantum efficiency measured at \SI{2.5}{\volt} (Sec.~\ref{sec:qe}) with the QEFR measured at \SI{10.5}{\volt} (Sec.~\ref{sec:qefr}). The value of \SI{10.5}{\volt} was chosen as a good compromise between sensitivity and operability, since it combines a high SNR (Sec.~\ref{sec:snr}) with a still limited fraction of outlier pixels (Sec.~\ref{sec:prnu}). Since the QEFR is the ratio $\mathrm{QE}/F$, inverting it gives
\begin{equation}
\label{eq:F}
F = \frac{\mathrm{QE}}{\mathrm{QEFR}} = \frac{0.46}{0.40} = 1.15 \pm 0.14\, .
\end{equation}
This determination relies on the assumption that the quantum efficiency is the same at \SI{10.5}{\volt} as at \SI{2.5}{\volt}, where it was measured. This assumption is required because the QEFR is measured in the avalanche regime while the quantum efficiency can only be measured below the avalanche onset, but it is almost certainly not exact: as discussed in Sec.~\ref{sec:qe}, Leonardo expects the quantum efficiency to keep increasing up to \SI{6}{\volt}. If the true quantum efficiency at \SI{10.5}{\volt} is higher than the value measured at \SI{2.5}{\volt}, then the excess noise factor is correspondingly higher than the value reported here. The value of \num{1.15} is therefore a lower estimate.

The measured value is consistent with the low excess noise expected from the single-carrier, electron-initiated avalanche of HgCdTe APDs, in which the multiplication is nearly deterministic~\cite{Rothman-2018}. For comparison, excess noise factors between \num{1.0} and \num{1.3} have been reported on Leonardo SAPHIRA SWIR APDs over a gain range from \num{1} to \num{400}~\cite{Finger-2023a}. At a gain of about \num{10}, comparable to that of IBEX at \SI{10.5}{\volt}, this reference gives an excess noise factor between \num{1.15} and \num{1.2}, in excellent agreement with our measurement. This suggests that the larger format of the IBEX detector does not degrade the quality of the avalanche process.

\section{CONCLUSION AND PERSPECTIVES}
\label{sec:conclusion}

We have reported the first electro-optical characterization of a Leonardo IBEX $2048\times2048$ HgCdTe LmAPD array, operated at \SI{80}{\kelvin} on a dedicated bench at CEA-IRFU. After establishing the signal and noise model of a linear-mode APD, we discussed the central difficulty of characterizing such a device, namely the degeneracy between the avalanche gain, the quantum efficiency, and the conversion gain in the measured response. To address it, we separated the measurements into two categories: the figures of merit that are directly measurable without assumptions, and those that require an explicit set of assumptions to be extracted individually.

Table~\ref{tab:summary} collects the measured figures of merit, giving a concise picture of what the community can expect from an IBEX detector. Among the directly measurable quantities, the SNR of a CDS measurement increases with pixel bias and exceeds that of a Euclid-type H2RG above \SI{8}{\volt} under identical low-flux conditions, establishing IBEX as a credible European alternative to HxRG detectors for photon-starved applications. The PRNU remains stable at the $\sim$10\% level, with fewer than 12\% outliers up to \SI{12.5}{\volt}. The QEFR, an assumption-free measurement of the information loss in the shot-noise regime obtained directly from the slope of the $\mathrm{SNR}^2$ curve, reaches \num{0.40} at \SI{10.5}{\volt}, a value close to that reported on single-element APDs and which we believe could serve as a reference metric for comparing the intrinsic quality of APD photosensitive layers. Among the derived quantities, the conversion gain corresponds to a sense-node capacitance of \SI{27}{\femto\farad}, in agreement with the \SI{27.3}{\femto\farad} expected by Leonardo. Under the assumption that the avalanche has not started at \SI{2.5}{\volt}, the quantum efficiency is $46\pm12\%$ at this bias, a value lower than expected that will be investigated on another IBEX detector. Assuming in addition that the quantum efficiency remains constant up to \SI{10.5}{\volt}, the excess noise factor is $F=1.15\pm0.14$ at that bias, consistent with the low excess noise of HgCdTe LmAPDs and with values reported on SAPHIRA arrays at comparable gain. This value should be regarded as a lower estimate, since a higher quantum efficiency at \SI{10.5}{\volt} would raise it. Taken together, these results establish IBEX as a promising European large-format detector for future ultra-low-flux NIR astronomy.

Several studies are already planned or ongoing to complete this characterization. An extensive study of the dark current and its evolution with temperature and pixel bias is ongoing and will be published shortly, along with studies of the crosstalk and the persistence and their evolution with pixel bias. The quantum efficiency measured on this particular IBEX detector is lower than expected, and a comparison with another IBEX detector is planned to confirm whether it originates from a defect in the sensitive layer. We are also developing methods that could allow the conversion gain to be measured independently and its precise evolution with pixel bias to be followed, which would help lift the degeneracy discussed in Sec.~\ref{sec:challenges}. All these performances will additionally be measured at \SI{40}{\kelvin} and \SI{120}{\kelvin} to characterize their dependence on the operating temperature. Finally, a proton-irradiation campaign is scheduled for October 2026 to assess the radiation hardness of the technology.

\begin{table}[ht]
\caption{Summary of the measured IBEX figures of merit.}
\label{tab:summary}
\begin{center}
\begin{tabular}{|l|l|}
\hline
\rule[-1ex]{0pt}{3.5ex} Figure of merit & Measured value \\
\hline
\rule[-1ex]{0pt}{3.5ex} Excess noise factor $F$ (\SI{10.5}{\volt}) & $1.15\pm0.14$ \\
\hline
\rule[-1ex]{0pt}{3.5ex} QEFR (\SI{10.5}{\volt}) & $0.40\pm0.05$ \\
\hline
\rule[-1ex]{0pt}{3.5ex} Quantum efficiency (\SI{2.5}{\volt}) & $46\pm12\%$ \\
\hline
\rule[-1ex]{0pt}{3.5ex} PRNU & $\sim$10\% \\
\hline
\rule[-1ex]{0pt}{3.5ex} Conversion gain & \SI{7.42}{\electron\per\ADU} \\
\hline
\rule[-1ex]{0pt}{3.5ex} Sense-node capacitance & \SI{27}{\femto\farad} (exp. \SI{27.3}{\femto\farad}) \\
\hline
\end{tabular}
\end{center}
\end{table}

\FloatBarrier
\appendix\section{DERIVATION OF THE SIGNAL VARIANCE}
\label{sec:variance-derivation}

We derive here the variance of the signal defined in Eq.~(\ref{eq:signal-sum}). Three assumptions are made. First, the multiplications undergone by the individual primary electrons are mutually independent. Second, the gain statistics are identical for all primary electrons and independent of the number of absorbed photons, so that $\langle M_i \rangle = \langle M \rangle$ and $\mathrm{Var}(M_i) = \sigma_M^2$ for all $i$. Third, the photo-generated primary electrons obey Poisson statistics, which follows from the Poisson statistics of the incident photons since a binomial selection of a Poisson process remains Poisson-distributed.

For conciseness we write $N \equiv N_{\mathrm{ph}} \times QE$ for the number of primary electrons, whose mean is $\langle N \rangle = QE \times \langle N_{\mathrm{ph}} \rangle$. The signal is then $S = \frac{1}{c_g}\sum_{i=1}^{N} M_i$, a sum with a random number of random terms. Its variance is obtained from the law of total variance,
\begin{equation}
\label{eq:app-total-variance}
\mathrm{Var}(Y) = \mathbb{E}_X\!\left[\mathrm{Var}(Y|X)\right] + \mathrm{Var}_X\!\left[\mathbb{E}(Y|X)\right] ,
\end{equation}
applied by conditioning on $N$,
\begin{equation}
\label{eq:app-var-decomp}
\sigma_S^2 = \frac{1}{c_g^2} \left[ \underbrace{\mathbb{E}_N\!\left(\mathrm{Var}\!\left(\textstyle\sum_{i=1}^{N} M_i \,\middle|\, N\right)\right)}_{\text{gain dispersion}} + \underbrace{\mathrm{Var}_N\!\left(\mathbb{E}\!\left(\textstyle\sum_{i=1}^{N} M_i \,\middle|\, N\right)\right)}_{\text{photon shot noise}} \right] .
\end{equation}

\subsection*{First term: gain dispersion}

At a fixed number $N$ of primary electrons, the sum contains a deterministic number of terms. Since the multiplications are mutually independent, the variances add,
\begin{equation}
\label{eq:app-first-cond}
\mathrm{Var}\!\left(\sum_{i=1}^{N} M_i \,\middle|\, N\right) = \sum_{i=1}^{N} \mathrm{Var}(M_i) = N\,\sigma_M^2 .
\end{equation}
From the definition of the excess noise factor in Eq.~(\ref{eq:enf}), $\sigma_M^2 = \langle M \rangle^2 (F-1)$. Taking the expectation over the photon statistics gives
\begin{equation}
\label{eq:app-first-term}
\mathbb{E}_N\!\left(N\,\sigma_M^2\right) = \langle N \rangle\,\langle M \rangle^2 (F-1) .
\end{equation}

\subsection*{Second term: photon shot noise}

At fixed $N$, each term contributes $\langle M \rangle$ on average, so that
\begin{equation}
\label{eq:app-second-cond}
\mathbb{E}\!\left(\sum_{i=1}^{N} M_i \,\middle|\, N\right) = N\,\langle M \rangle .
\end{equation}
Since $\langle M \rangle$ is a constant, the variance of this quantity over the photon statistics is
\begin{equation}
\label{eq:app-second-term}
\mathrm{Var}_N\!\left(N\,\langle M \rangle\right) = \langle M \rangle^2\,\mathrm{Var}(N) = \langle N \rangle\,\langle M \rangle^2 ,
\end{equation}

\subsection*{Result}

Substituting Eq.~(\ref{eq:app-first-term}) and Eq.~(\ref{eq:app-second-term}) into Eq.~(\ref{eq:app-var-decomp}) gives
\begin{equation}
\label{eq:app-sum}
\sigma_S^2 = \frac{1}{c_g^2} \left[ \langle N \rangle\,\langle M \rangle^2 (F-1) + \langle N \rangle\,\langle M \rangle^2 \right] = \frac{\langle N \rangle\,F\,\langle M \rangle^2}{c_g^2} ,
\end{equation}
that is, using the explicit notation,
\begin{equation}
\label{eq:app-var-signal}
\sigma_S^2 = \frac{QE \times \langle N_{\mathrm{ph}} \rangle \times F \times \langle M \rangle^2}{c_g^2}\, ,
\end{equation}

\acknowledgments

This work was carried out at the D\'epartement d'Astrophysique (DAP) of CEA-IRFU and was co-funded by the European Space Agency (ESA). The authors thank Leonardo UK for the IBEX detector and for helpful discussions, and thank their colleagues at CEA and ESA involved in the IBEX programme.

\bibliography{report}
\bibliographystyle{spiebib}

\end{document}